\documentclass[pra,twocolumn,superscriptaddress]{revtex4}

\usepackage{amsmath,amssymb,mathrsfs}
\usepackage{subfigure}
\usepackage[dvips]{graphicx}
\usepackage{hyperref}
\usepackage{paralist}
\usepackage{epstopdf}
\usepackage{dcolumn}
\usepackage{color}

\def\be{\begin{equation}}
\def\ee{\end{equation}}
\def\bea{\begin{eqnarray}}
\def\eea{\end{eqnarray}}
\def\nn{\nonumber}

\def\psibar{{\bar{\psi} }}

\def\Jbar{{\bar{J}}}

\begin{document}

\title{Orbital coupled dipolar fermions in an asymmetric optical ladder}

 \author{Xiaopeng Li}
\affiliation{Department of Physics and Astronomy, University of Pittsburgh, Pittsburgh, Pennsylvania}
\affiliation{Kavli Institute for Theoretical Physics, University of California, Santa Barbara, CA 93106, USA}

\author{W. Vincent Liu\footnote{email: w.vincent.liu@gmail.com}}
\affiliation{Department of Physics and Astronomy, University of Pittsburgh, Pittsburgh, Pennsylvania}

\begin{abstract}
We study a quantum ladder of interacting fermions with coupled $s$
and $p$ orbitals. Such a model describes dipolar molecules or atoms loaded
into a double-well optical lattice, dipole moments being aligned by an
external field. The two orbital components have distinct hoppings.
The tunneling between them is equivalent to a partial Rashba spin-orbital coupling
when the orbital space ($s$, $p$) is identified as spanned by pseudo-spin 1/2 states.
A rich
phase diagram, including incommensurate orbital density wave, pair density wave
and other exotic superconducting phases, is proposed with
bosonization analysis.  In particular, superconductivity is found in
the repulsive regime.
\end{abstract}

\maketitle

\section{Introduction}
Orbital degree of freedom~\cite{2011_Lewenstein_Liu_orbital_dance} plays a fundamental
role in understanding the unconventional
properties in solid state materials~\cite{2000_Tokura_Nagaosa_orbital_Science}.
Recent experiments in optical lattices
have demonstrated that orbitals  can also be used to construct quantum emulators of
exotic models beyond natural crystals. Orbital lattices are attracting growing interests
due to their unique and fascinating properties resulting
from the spatial nature of the degenerate states.
For example, the bosonic $p_x + ip_y$
superfluid~\cite{2005_Isacsson_pband,2006_Liu_TSOC,2006_Kuklov_sf,2008_Lim_TSOC,2012_Li_1DTSOC_PRL}
state has been prepared on a bipartite square lattice~\cite{2011_Wirth_pband}, and later
the other complex superfluid
with $s$ and $p$ orbitals correlated was observed on a hexagonal lattice~\cite{2011_Sengstock_honeycomb_BEC}.

Previous study on multicomponent cold gases mainly focused on hyperfine states of
alkali atoms~\cite{2008_Bloch_review,2012_Lewenstein_OLEreview}. In a cold gas of atoms
with two approximately degenerate
hyperfine states,  the realized pseudo-spin SU(2) symmetry makes it possible to emulate
Fermi Hubbard model in optical lattices~\cite{2006_Bloch_fermi,2007_Scalapino_FHEmulator,2009_Giamarchi_FHEmulator}.
To engineer spin-orbital couplings and the resulting topological phases, one has to induce Raman transitions between
the hyperfine states to break the pseudo-spin
symmetry~\cite{2009_Liu_SOC_PRL,2009_Spielman_SOCBoson,2012_Wang_JingZhang_SOCFermi,2012_Zwierlein_SOCFermi,2012_Zhang_Pan_SOC_PRL}.
In contrast,
due to the spatial nature of the orbital degrees of freedom, the
symmetry in orbital gases, such as that in $p_x + ip_y$
superfluid~\cite{2011_Wirth_pband}, can be controlled by simply changing the
lattice geometry
{as shown in Ref.~\cite{2011_Wirth_pband,2011_Sengstock_honeycomb_BEC,2012_Esslinger_megeDirac_Nature,2012_Hemmerich_TopAvoidCross_PRL},
where unprecedented tunability of double-wells has been demonstrated.}
With a certain lattice geometry, a spin-orbital like coupling
can naturally appear in an orbital gas with $s$ and $p$-orbitals without
Raman transitions~\cite{2012_Li_orbitalladder_NatComm}. Theoretical studies of orbital physics largely focusing
on two or three dimensions suggest exotic orbital
phases~\cite{2005_Isacsson_pband,2006_Liu_TSOC,2006_Kuklov_sf,2008_Zhao_pOrborder,2008_Wu_pOrbOrder,2008_Lim_TSOC,2008_Vladimir_icsf,2009_Wu_pband,2009_Hung_fpair,2010_Cai_FFLO,2010_Zhou_interband,2011_Sun_TSM,2012_DasSarma_dwBEC_PRA}
beyond the scope of spin physics.

In this article, we study a one dimensional orbital ladder with $s$ and $p$ orbitals
coupled~\cite{2010_Zhang_sppair,2012_Li_orbitalladder_NatComm}. We shall
derive such an effective model for dipolar
molecules or atoms~\cite{2008_JILA_Dip_Science,2010_Benjamin_Dysprosium_PRL,2011_JILA_Dip_NPHYS,2012_JILA_Dip_PRL,2012_Zwierlein_polar_PRL}
loaded in a
double-well optical lattice.
The tunneling rates (or effective mass) of each orbital component are highly tunable by
changing the lattice strength.
The coupling between $s$ and $p$ orbitals mimics the spin-orbital couplings~\cite{2012_Li_orbitalladder_NatComm}.
This orbital system suggests the possibility of exploring the equivalent of the exciting
spin-orbital coupled physics in dipolar gases yet without requiring the use of synthetic gauge
fields, and hence it provides an interesting, simple alternative route.
{A rich phase diagram, including incommensurate orbital density wave (ODW),
pair density wave (PDW)~\cite{2009_Berg_Fradkin_PDW_NPHYS,2012_Jaefari_Fradkin_PDW_PRB,2012_Wei_PDW_PRL,2012_Robinson_PDW_PRB},
 and other exotic superconducting phases, is found
with bosonization analysis.  The PDW phase realized here
is a superconducting phase, that features an oscillating Cooper pair field
with a period of $\pi$.}
The incommensurate ODW phase has an oscillating particle-hole
pair, which tends to break
the time-reversal symmetry. An exotic superconducting phase on the repulsive side is also discovered.


\begin{figure}[htp]
\begin{center}
\includegraphics[angle=0,width=.9\linewidth]{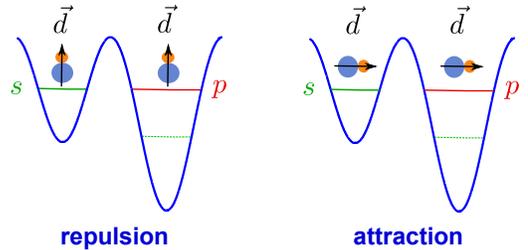}
\end{center}
\caption{{ (Color online) Schematic plot  illustrating control of interactions 
with polar molecules or atoms loaded on $s$- and $p$- orbitals. Dipole moments aligned 
``head-to-head'' (``head-to-tail'') in the left (right) figure provide repulsive (attractive) interaction.  }
}
\label{fig:dipoleschematic}
\end{figure}

\section{Model}
 Consider a cold ensemble of polar molecules or atoms,
e.g. $^{40}\mathrm{K}^{87}\mathrm{Rb}$~\cite{2008_JILA_Dip_Science,2011_JILA_Dip_NPHYS,2012_JILA_Dip_PRL}, 
$^{23}\mathrm{Na}^{40}\mathrm{K}$~\cite{2012_Zwierlein_polar_PRL}, 
$\mathrm{OH}$~\cite{2012_JILA_OH_Nature}, 
or Dy~\cite{2010_Benjamin_Dysprosium_PRL},  whose dipole moments are
controlled by an external field as demonstrated in
experiments.
Long lived polar molecules have been realized in optical lattices~\cite{2012_JILA_Dip_PRL}.
Let the ensemble trapped by a ladder-like optical
lattice of the type studied
in~\cite{2012_Li_orbitalladder_NatComm}.
{As shown in the schematic picture in~\cite{2012_Li_orbitalladder_NatComm},
the lattice consists
of two chains of potentials of unequal depth. 
We consider a single species of fermionic atoms/molecules occupying the
$s$ and $p$ orbitals of the shallow and deep chains, respectively, 
with the low-lying $s$ orbitals on the deep chain completely filled (FIG.~\ref{fig:dipoleschematic}).  
Alternatively, fermions can be directly 
loaded into the higher orbitals while keeping the lower $s$ nearly empty
 by the techniques developed in recent
experiments~\cite{2011_Wirth_pband,2012_Esslinger_megeDirac_Nature}. 
Coherent meta-stable states in high orbitals 
with long life time up to several hundred milliseconds were demonstrated achievable~\cite{2011_Wirth_pband}.
To suppress chemical reactions of polar molecules, the latter approach is preferable. }
The single particle Hamiltonian of the $sp$-orbital ladder is then
given as~\cite{2012_Li_orbitalladder_NatComm}
\bea
H_0 &=& \sum_j C_j ^\dag
  \left[\begin{array}{cc}
         -t_s & -t_{sp} \\
	  t_{sp} & t_p
        \end{array}
   \right]
  C_{j+1} +h.c. 
\eea
where $C_j ^\dag = [a_s ^\dag (j), a_p ^\dag (j) ]$, and $a_s ^\dag $
($a_p ^\dag$) is
the creation operator for the $s$-orbital ($p$-orbital). The lattice constant is set as the
length unit in this paper.
{In the proposed optical lattice setup~\cite{2012_Li_orbitalladder_NatComm}, the ratios
$t_s/t_p$ and $t_{sp}/t_p$ are small (typically $0.1$). 
{
We emphasize here that $s$-orbital is parity even and that $p$-orbital is parity odd. 
The relative signs of hoppings are dictated by the parity nature of the $s$- and $p$-orbitals~\cite{2012_Li_orbitalladder_NatComm}}

The band structure is readily obtained by Fourier transform
$C_j  = \int \frac{dk}{2\pi} \tilde{C} (k) e^{i k j}$. The Hamiltonian in the momentum
space reads as $H_0 = \int \frac{dk}{2\pi} \tilde{C} ^\dag (k) \tilde{\cal H} (k) \tilde{C} (k)$, with
$
 \tilde{ \mathcal{H}}  (k) = h_0 (k) \sigma_0 + \vec{h} (k) \cdot \vec{\sigma},
$
where $h_0 (k) = (t_p - t_s ) \cos(k)$, $h_x (k) =0$, $h_y (k)  = 2 t_{sp} \sin(k)$ and
$h_z (k ) = -(t_p + t_s) \cos (k)$.
Here $\sigma_0$ is the identity matrix and $\sigma_{x,y,z}$ are Pauli matrices.
The two bands
are given by
$
E_\pm (k) = h_0 (k) \pm
    \sqrt{ h_y ^2 (k) + h_z^2 (k) },
$
which are shown in FIG.~\ref{fig:spectra}.
The Hamiltonian is rewritten as
$H_0 = \int \frac{dk}{2\pi} \sum_{\wp=\pm}  E_{\wp} (k) \phi_{\wp} ^\dag(k) \phi_{\wp} (k)$.
We define an angle variable $\theta$ by
$\cos(\theta (k)) = h_z/|\vec{h}|$ and $\sin(\theta (k)) = h_y/|\vec{h}|$ to save writing.
Here, we only consider lower than half filling, i.e., less than one particle per unit cell. The lower band
is thus partially filled and the upper band is empty. Since we are interested in the low-energy physics, the spectrum
$E_-$ is linearized around the Fermi momenta $\pm k_{F\nu}$. Here, $\nu = A$ or $B$, and  $\pm k_{FA}$ are
inner Fermi points and $\pm k_{FB}$ are outer Fermi points (FIG.~\ref{fig:spectra}).
The resulting Fermi velocities are
$v_{F\nu} = |\frac{\partial E_- (k)}{\partial k }|_{k=k_{F\nu}} $.
The operators capturing the
low energy fluctuations are defined with
right ($\Psi$) and left ($\overline{\Psi}$)  moving modes
$\Psi_A (k) = \phi_- (k_{FA} +k)$, $\overline{\Psi}_A (k) = \phi_- (-k_{FA} +k)$,
$\Psi_B (k) = \phi_- (-k_{FB} +k)$ and $\overline{\Psi}_B (k) = \phi_- (k_{FB} +k)$.
The field operators
are introduced by
$\psi_\nu (x) = \int \frac{dk}{2\pi} \Psi_\nu (k) e^{ikx}$ and
$\psibar_\nu (x) = \int \frac{dk}{2\pi} \overline{\Psi}_\nu (k) e^{ikx}$.
These field operators are related to lattice operators by
\bea
C(j)  &\to &\lambda ^{A} \psi_A (x) e^{ik_{FA} x} + \lambda ^{A*} \bar{\psi}_A (x) e^{-ik_{FA} x} \nn \\
&+ &\lambda ^B \psi_B (x) e^{-ik_{FB} x} + \lambda ^{B*} \bar{\psi}_B (x) e^{i k_{FB} x},
\label{eq:Cj}
\eea
where
\[\lambda ^\nu =
\left[ \begin{array}{c}
 i \sin(\theta_\nu/2) \\
 \cos(\theta_\nu/2)
\end{array}\right],
\]
with $\theta_A = \theta (k_{FA})$ and $\theta_B = \theta (-k_{FB})$.
{The substitution in Eq.~(\ref{eq:Cj}) and the energy linearization are valid
for weakly interacting fermions at low temperature. }

With polar molecules or atoms loaded on the $sp$-ladder, we include all
momentum-independent interactions (momentum-dependent part is
irrelevant in the Renormalization group flow~\cite{1994_Shankar_RGFermion_RMP})
allowed by symmetry. The Hamiltonian density of the interactions is given by
\bea
\mathcal{H}_{\text{int} }
&=& \sum_{\nu \nu'}  \frac{1}{2}g_4 ^{\nu \nu'} \left[ J_\nu J_{\nu'} + \Jbar_\nu \Jbar_{\nu'} \right]
 + g_2 ^{\nu \nu'} J_\nu \Jbar_{\nu'} \nonumber \\
&& + g_3 \left\{\psibar_A ^* \psibar_B \psi^* _A \psi_B  +
		\psibar_B ^* \psibar_A \psi^* _B \psi_A \right\},
\eea
where $J_\nu = :\psi_\nu ^* \psi_\nu :$ and $\Jbar_\nu = :\psibar _\nu ^* \psibar_\nu:$. For
the symmetric case $t_s = t_p$, an Umklapp process
\be
 \mathcal{H}_{um} = g_u\left\{ \psibar_A ^* \psi{_A} \psi_B ^* \psibar{_B} +
			      \psibar_B ^* \psi{_B} \psi_A ^* \psibar{_A} \right\}
\ee
becomes allowed for the reason that $k_{FA} + { k_{FB} } = \pi$.
Since dipolar interactions between polar
molecules or atoms decay as $1/r^3$,
the leading interaction in the proposed double-well lattice
setup~\cite{2012_Li_orbitalladder_NatComm} is
$$
 H_{\text{int}} = U \sum_j
	\left[a_s ^\dag (j) a_s (j) -\frac{1}{2} \right]
	\left[a_p ^\dag (j) a_p (j) - \frac{1}{2} \right].
$$
The strength of $U$ is tunable by changing the dipole moment, 
{ or by 
varying the distance between the shallow and deep wells (FIG.~\ref{fig:dipoleschematic}). 
By controlling this distance the leading interaction can be made significantly 
larger than sub-leading interactions (dipolar tails), which are neglected here.  
}

In the weak interacting limit,
the $g$-ology couplings are related to $U$ by
$g_4 ^{\nu \nu} = U$, $g_4 ^{AB} = g_4 ^{BA} = U \sin^2 \left(\frac{\theta_A - \theta_B}{2}\right)$,
$g_2 ^{\nu \nu} = U \sin^2 (\theta_\nu)$, $g_2 { ^{AB} } = g_2 { ^{BA} } = U$,
\[
g_3 = U \sin(\theta{_A}) \sin(\theta{_B}) ,\]
and
\[ g_u = U \cos(\theta{_A}) \cos(\theta{_B}), \]
at tree level~\cite{1994_Shankar_RGFermion_RMP}.
Considering strong interactions or the finite ranged tail of
dipolar interactions, the g-ology couplings will be renormalized
due to neglected irrelevant couplings.
By manipulating the direction of dipole moments with {an external field,}
the interaction can
be either repulsive or attractive 
(FIG.~\ref{fig:dipoleschematic})~\cite{2008_JILA_Dip_Science,2011_JILA_Dip_NPHYS,2012_JILA_Dip_PRL,2012_Zwierlein_polar_PRL}.

We follow the notation convention of Ref.~\cite{1999_Senechal_bosonization_arXiv}, where
the bosonization identity takes the form
\bea
\psi_\nu = \frac{1}{\sqrt{2\pi} } \eta_\nu e^{-i\sqrt{\pi} (\varphi_\nu + \vartheta_\nu) } \nonumber \\
\psibar_\nu = \frac{1}{\sqrt{2\pi}} \bar{\eta}_\nu e^{i\sqrt{\pi} (\varphi_\nu - \vartheta_\nu) },
\eea
where $\eta_\nu$ is the Klein factor and $\vartheta_\nu$ is the dual field of boson field $\varphi_\nu$.
The charge and orbital boson fields are further introduced here by
$[\varphi_c, \varphi_o] = [\varphi{_A}, \varphi{_B}] T$, with the matrix $T$ given by
$$
T = \frac{1}{\sqrt{2}} \left[
\begin{array}{cc}
 1& - 1\\
  1 & 1
\end{array}
\right] .
$$
and their dual fields are $[\vartheta_c , \vartheta_o] = [\vartheta{_A}, \vartheta{_B}] T$.
The Bosonized Hamiltonian density reads
\bea
{\cal H} &=& {\cal H}_c + {\cal H}_o + {\cal H}_\text{mix} ,\nonumber  \\
{\cal H} _c &=& \frac{u_{c}}{2}  \left[ K_c \Pi_{c} ^2 + \frac{1}{K_c} (\partial_x \varphi_c)^2 \right],   \nonumber \\
{\cal H}_o &=& \frac{u_{{o}}}{2}  \left[ K_o \Pi_{{o}} ^2 + \frac{1}{K_o} (\partial_x \varphi_o)^2 \right] \nonumber \\
	&& +\frac{1}{2\pi^2}  \left[ {g_3} \cos (\sqrt{8\pi} \vartheta_o)
			  +{g_u} \cos(\sqrt{8\pi} \varphi_o)\right],  \nonumber \\
{\cal H}_\text{mix} &=& u_{m}   \left [ K_{m} \Pi_c \Pi_o
	+ \frac{1}{K_{m}} (\partial_x \varphi_c) (\partial_x \varphi_o) \right],
\eea
with
$$
u_{\alpha=c/o }  =
    \sqrt{\left(v_+ +\tilde{g}_4^{\alpha\alpha } /2\pi\right)^2 - \left(\tilde{g}_2 ^{\alpha\alpha } /2\pi\right) ^2} ,$$
$$
K_{\alpha} =
\sqrt{ \frac{2\pi v_+ +\tilde{g}_4 ^{\alpha \alpha } -\tilde{g}_2 ^{\alpha \alpha} }
	    {2\pi v_+ +\tilde{g}_4 ^{\alpha \alpha } +\tilde{g}_2 ^{\alpha \alpha}} },
$$
$$
 u_{m} = \sqrt{\left(v_- +\tilde{g}_4^{c o} /2\pi\right)^2 - \left(\tilde{g}_2 ^{c o } /2\pi\right) ^2},
$$ and
$$
 K_m =
\sqrt{ \frac{2\pi v_- +\tilde{g}_4 ^{c o} -\tilde{g}_2 ^{c o} }
	    {2\pi v_- +\tilde{g}_4 ^{c o} +\tilde{g}_2 ^{c o}} },
$$
where $v_+ = ({ v_{FA} } + { v_{FB} })/2$, $v_- = (-{ v_{FA} } + { v_{FB} })/2$ and the transformed coupling matrices $\tilde{g}_4$ and
$\tilde{g}_2$ are given by
$[\tilde{g}] = T^{-1} [g] T$.
The mixing term ${\cal H}_\mathrm{mix}$ vanishes for the symmetric case with $t_s = t_p$.

\begin{figure}[htp]
\begin{center}
\includegraphics[angle=0,width=1.0\linewidth]{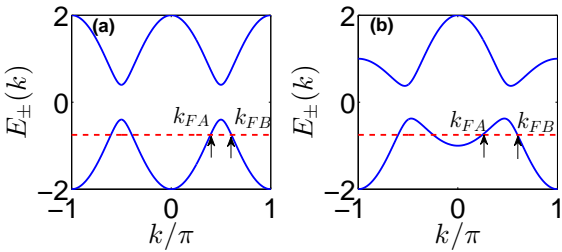}
\vspace{-3em}
\end{center}
\caption{(Color online) Sketch of the band structure of the $sp$-orbital ladder. Red dashed lines show the level of chemical potentials.
(a), the symmetric case with $t_s = t_p$. (b), the asymmetric case with $t_s <t_p$. { Here $t_p$ is taken as the energy unit.} 
}
\label{fig:spectra}
\end{figure}

\section{Quantum phases and transitions of the Symmetric case}
For the symmetric case with
$t_s = t_p$ (FIG.~\ref{fig:spectra}), the Hamiltonian has
an accidental $Z_2$ symmetry, $C_j \to (-1)^j \sigma_x C_j$ and Fermi momenta are
related by $k_{FA} = \pi - { k_{FB} } \equiv k_F$.
This $Z_2$ symmetry implies that
${ v_{FA} } = { v_{FB} }$, $g_4 { ^{AA} } = g_4 { ^{BB} }$ and $g_2 { ^{AA} } = g_2 { ^{BB} }$.
We find that the transformed
coupling matrices $\tilde{g}_2$ and $\tilde{g}_4$ are diagonal and that
the orbital-charge mixing term ${\cal H}_\mathrm{mix}$ vanishes.
In other words, the $Z_2$ symmetry guarantees orbital-charge separation.
The charge part $H_c$ is quadratic and the orbital part $H_o$
is a sine-Gordon model~\cite{2003_Giamarchi_oned}.

With attraction, we have $K_o<1$, $g_u>0$, and
the sine-Gordon term $g_u$ is relevant (flows to $+\infty$) in the renormalization group (RG) flow~\cite{2003_Giamarchi_oned}.
This corresponds to an orbital gapped phase with $\cos(\sqrt{8\pi} \varphi_o)$ locked at $-1$. In this phase, quantum fluctuations
of $\varphi_o$ become massive, and the divergent susceptibilities are the following: charge density wave (CDW)  and
PDW~\cite{2009_Berg_Fradkin_PDW_NPHYS,2012_Jaefari_Fradkin_PDW_PRB,2012_Robinson_PDW_PRB}
given by the operators:
\bea
   O_\text{CDW} (x) &=& \psi_A ^* \psibar{_A} e^{-2ik_{FA} x} - \psi_B ^* \psibar{_B} e^{2i{ k_{FB} } x}   \nonumber \\
	   &\propto& e^{-2ik_{F} x} e^{i\sqrt{2\pi} \varphi_c } \sin (\sqrt{2\pi} \varphi_o) \nonumber \\
   O_\text{PDW} (x) &=& \psi{_A} \psibar{_B} e^{i(k_{FA}+{ k_{FB} }) x} + \psi{_B} \psibar{_A} e^{-i ({ k_{FA} } + { k_{FB} }) x} \nonumber \\
	 &\propto& (-1)^x e^{-i\sqrt{2\pi} \vartheta_c} \sin (\sqrt{2\pi} \varphi_o) \nonumber
\eea
Due to orbital-charge separation, the CDW and PDW correlation functions are readily given by
\bea
\langle O_\text{CDW} (x) O_\text{CDW} ^\dag (0) \rangle &\propto&  e^{-2ik_{F} x} x^{-K_c}, \\
\langle O_\text{PDW} (x) O_\text{PDW} ^\dag (0)\rangle &\propto & (-1)^x x^{-1/K_c}.
\label{eq:symmetric_pdw}
\eea
Since $K_c >1$ for attraction, the algebraic PDW order is dominant. In this phase,
the superconducting pairing $\mathcal{O}_\text{SC} = a_s (j) a_p (j)$ oscillates in space
with a period of $\pi$.

With repulsion, we have $K_o >1$, and thus $g_3$ is relevant~\cite{2003_Giamarchi_oned}.
This gives an orbital gapped phase with $\cos(\sqrt{8\pi} \vartheta_o)$
locked at $1$, because $g_3<0$. The fluctuations of $\vartheta_o$ are massive, and the divergent susceptibilities are
ODW and  superconducting SC$^+$ given by the operators:
\bea
   O_\text{ODW} (x) & =& e^{-i({ k_{FA} } - { k_{FB} }) x} ( \psi_A ^* \psibar {_B}
			- \psi_B ^* \psibar{_A}  )  \nonumber \\
	&\propto& e^{-i({ k_{FA} } - { k_{FB} }) x} e^{i\sqrt{2\pi} \varphi_c} \cos (\sqrt{2\pi} \vartheta_o) \nonumber \\
   O_{\text{SC}^+} (x) &=& \psi{_A} \psibar{_A} +  \psi{_B} \psibar{_B} \nonumber \\
	&\propto & e^{-i\sqrt{2\pi} \vartheta_c} \cos (\sqrt{2\pi} \vartheta_o) \nonumber
\eea
Since $K_c <1$ for repulsion, the dominant algebraic order here is ODW, for which the correlation function is given by
\be
 \langle O_\text{ODW} (x) O_\text{ODW} ^\dag (0)\rangle \propto e^{-i({ k_{FA} } - { k_{FB} }) x} x^{-K_c}.
\label{eq:symmetric_oaf}
\ee
In the ODW phase, the particle-hole pairing in terms of lattice operators reads
$\mathcal{O}_\text{ODW} (j) =   C_j ^\dag \sigma_y C_j$.
This ODW order is incommensurate with an oscillation period $2\pi/({ k_{FA} } - { k_{FB} })$ in real space.
If we go beyond the one-dimensional limit and consider small transverse
tunnelings~\cite{2012_Li_orbitalladder_NatComm}, a true long-range
ODW order $\langle O_\mathrm{ODW} (x) \rangle \propto e^{i({ k_{FA} } - { k_{FB} }) x}$ is expected. Such an order
breaks time-reversal symmetry.

The ODW and PDW phases predicted by Bosonization analysis are further verified in numerical simulations
with matrix products state,
{ in which open boundary condition is adopted.}
The superconducting correlation
$$
C_\mathrm{SC} (j'-j) = \langle a_p ^\dag (j) a_s ^\dag (j) a_s (j') a_p (j') \rangle$$ 
and the orbital density wave
correlation 
$$
C_\mathrm{ODW} (j'-j) = \langle C_j ^\dag \sigma_y C_j C_{j'} ^\dag \sigma_y C_{j'} \rangle$$ 
are calculated.
{ In our calculation, the two points $j$ and $j'$ are $10$ sites away from the boundaries to
minimize the boundary effects.
The convergence of these correlations is checked in numerical simulations.}
FIG.~\ref{fig:phasediag_symmetric} shows the Fourier transform of these correlations, defined by
${\cal C}(k) =  \sum_{j\neq 0} C(j) e^{-i k j} $,
{which approaches to its thermodynamic limit with increasing system size (FIG.~\ref{fig:phasediag_symmetric}).}
The sharp peaks of ${\cal C}_\mathrm{SC}(k)$ at momenta $\pm \pi$
on the attractive side tell the quantum state has a PDW order shown in Eq.~\ref{eq:symmetric_pdw}.
{On the repulsive} side sharp dips of ${\cal C}_\mathrm{ODW} (k)$ at finite momenta verify the incommensurate ODW order
shown in Eq.~\ref{eq:symmetric_oaf}.
With numerical calculations, we also find the existence of PDW phase in the strongly attractive regime
if $t_s \neq t_p$.

\begin{figure}[htp]
\begin{center}
\includegraphics[angle=0,width=0.9\linewidth]{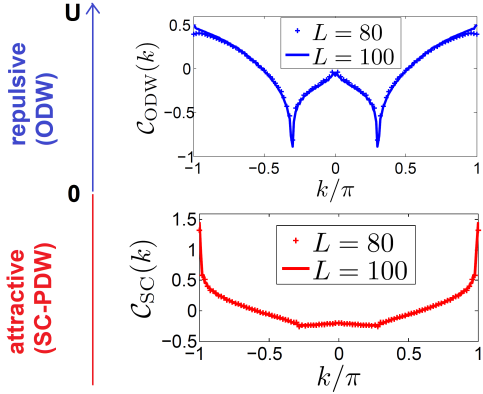}
\end{center}
\caption{(Color online) SC-PDW and CDW phases of the symmetric $sp$-orbital 
ladder with $t_s = t_p = 2t_{sp}$. ${\cal C}_\mathrm{ODW} (k)$ and
${\cal C}_\mathrm{SC} (k)$ show the Fourier transform of the orbital density wave and superconducting
correlations, respectively. {The numerical results are calculated 
for the system at two different sizes $L=80$ and $100$
at filling
$\frac{1}{L} \sum_j \langle a_s ^\dag (j) a_s(j) + a_p^\dag (j) a_p (j)\rangle = 0.7$.}  
In the upper (lower) graph, the interaction $U=3 t_s$ ($U =-3t_s$). }
\label{fig:phasediag_symmetric}
\end{figure}

\section{Quantum phases and transitions of the Asymmetric case}
For the asymmetric case---$t_s<t_p$ (FIG.~\ref{fig:spectra}), the Fermi velocity ${ v_{FB} } > {  v_{FA}  }$ and the orbital-charge separation no longer holds. Thus,
the orbital and charge degrees of freedom cannot be treated separately. The other difference with the symmetric case is
{that} the Umklapp process $g_u$ does not exist.
Since the effects of $g_4$ couplings are just to renormalize
the Fermi
velocities~\cite{1993_Fabrizio_coupledchain_PRB,1996_Balents_ladder_PRB,1997_Lin_Balents_Nchain_PRB,2000_Ledermann_twobandfermi_PRB}.
For simplicity, we do not consider such effects and set $g_4 ^{\nu \nu'} =0$ here.
The one-loop RG equations are given by~\cite{2000_Ledermann_twobandfermi_PRB},
\bea
\frac{d g_2 ^{\nu \nu'}} {dl} &=& \frac{g_3 ^2}{2\pi}
      \left[ \frac{ \delta _{\bar{\nu} \nu'} }{v_+}
      -\frac{\delta_{\nu \nu'}}{  v_{F\bar{\nu}} }
      \right], \nonumber \\
\frac{d g_3}{dl} &=& \frac{g_3}{2\pi}  \sum_\nu
      \left[ \frac{g_2 ^{\nu \bar{\nu}}}{ v_+}
	    -\frac{g_2^{\nu \nu}}{v_{F\nu}}
      \right],
\eea
where { $l$ is the flow parameter ($l \to \infty$) and  $\bar{\nu}=$ $A$ ($B$) for $\nu=$ $B$ ($A$).}
The RG flow of the sine-Gordon term $g_3$ is obtained as
\bea
&& \sqrt{|C|}/g_3 (l) \nonumber \\
& =&   F\left[-\text{sgn} (g_3 Y )\sqrt{\frac{2|C|D}{\pi v_+}} l +F^{-1} [\frac{\sqrt{|C|}}{ g_3 (0)}] \right],
\eea
with
\be
C = \frac{2{ v_{FA} } { v_{FB} } v_+^2 }{{ v_{FA} } { v_{FB} } + v_+^2 }
    \left[ \frac{g_2 { ^{AB} }}{v_+} - \frac{g_2 { ^{AA} }}{2{ v_{FA} }} - \frac{g_2 { ^{BB} }}{ 2 { v_{FB} }} \right]^2 -g_3 ^2,
\ee
\be
D = \frac{{ v_{FA} } { v_{FB} } +v_+ ^2}{\pi { v_{FA} } { v_{FB} } v_+},
\ee
and
\be
Y = \frac{g_2 { ^{AB} } }{v_+} -\frac{g_2 { ^{AA} }}{2 { v_{FA} }} - \frac{g_2 { ^{BB} }}{2 { v_{FB} }}.
\ee
The function $F$ is the hyperbolic function ``$\sinh$'' (the trigonometric function ``$\sin$'') if $C>0$ ($C<0$).
When $Y>0$, $g_3$ always flows to $\infty$ and the system is in some gapped phase. When $Y<0$, $g_3$ flows to $\infty$ only if
$C<0$.  
 $g_3$ is irrelevant only if
$
 C>0
$
and
$
 Y<0.
$
In the weak interacting regime,
we have 
\bea
Y/U = \frac{1}{v_+} -\frac{\sin^2 (\theta{_A})}{2 { v_{FA} }} - \frac{\sin^2 (\theta{_B})}{2 { v_{FB} } } .
\eea
We will consider the regime $Y/U>0$ { (this condition holds when $t_{sp}$ is weak compared with $t_s + t_p$)} in the following.

With repulsion ($U>0$, $Y>0$), $g_3$ is relevant and flows to $-\infty$ in RG flow.
Then the dual orbital field $\vartheta_o$ is locked with
$\cos(\sqrt{8\pi} \vartheta_o) = 1$ and its fluctuations $\vartheta_o$
are massive.
The key effect of orbital-charge mixing can be seen from its modification of the dynamics of the conjugate fields, given as
\bea
\Pi_{\theta_o} &=& \frac{K_o}{u_o} \partial_t \vartheta_o + \frac{K_o u_m}{K_m u_o} \partial_x \varphi_o, \\
\Pi_{\varphi_c} &=& \frac{1}{u_c K_c} \partial_t \varphi_c +\frac{K_m u_m}{K_c u_c} \partial_x \vartheta_o,
\eea
where $\Pi_{\varpi}$ is the conjugate field of $\varpi$. The Lagrangian is constructed by
$$\mathcal{L} (x,t) = \Pi_{\vartheta_o} \partial_t \vartheta_o + \Pi_{\varphi_c} \partial_t \varphi_c -\mathcal{H}.$$
With massive fluctuations of $\vartheta_o$ integrated out, the Lagrangian of the charge field $\varphi_c$ is
given by
\bea
\mathcal{L}_c = \frac{1}{2\gamma} \left[ \frac{1}{u} (\partial_t \varphi_c)^2 - u (\partial_x \varphi_c)^2 \right]
	+O\left((\partial \varphi_c)^4\right),
\eea
with the renormalized Luttinger parameter and sound velocity given by
\bea
 \gamma &=& \frac{K_c }{\sqrt{1-\frac{K_c K_o }{K_m^2}\frac{u_m^2 }{u_c u_o}} },  \\
 u &=& \sqrt{u_c ^2 - u_m^2 \frac{u_c K_c K_o}{u_o K_m^2}}.
\eea
To zeroth order in the interaction $U$, the renormalized Luttinger parameter is
\bea
\gamma = \left[{1 - \left(\frac{v_-}{v_+}\right)^2 }\right]^{-1/2}.
\label{eq:gamma}
\eea
Our result reproduces
the perturbative result~\cite{2000_Ledermann_twobandfermi_PRB} when the orbital-charge mixing term is small.
The diverging susceptibilities are ODW and SC$^+$, and the corresponding correlation functions
are given as
\bea
\langle  O_{\text{SC}^+}  (x) O_{\text{SC}^+} ^\dag  (0) \rangle &\propto&  x^{-1/\gamma}, \\
 \langle O_\text{ODW} (x) O_\text{ODW} ^\dag (0) \rangle &\propto&  e^{-i({ k_{FA} } - { k_{FB} }) x} x^{-\gamma}.
\eea
With sufficiently weak repulsion $\gamma >1$, the dominant order is SC$^+$, of which the pairing
in terms of lattice operators is $\mathcal{O}_\mathrm{SC} = a_s (j) a_p (j)$. We emphasize
here that this pairing does not oscillate in real space. 
{ Such a superconducting phase arising 
in the repulsive regime 
results from that charge mode $\varphi_c$ is coupled with the orbital mode $\varphi_o$, which 
is strongly fluctuating with its conjugate field $\vartheta_o$ pinned. }
 The sine-Gordon term $g_3$ causing this pinning effect is finite only when the coupling of $sp$-orbitals
$t_{sp}$ is finite, and $g_3$ is monotonically increasing when $t_{sp}$ is increased. Thus the transition temperature of
this repulsive superconducting phase can be increased by tuning $t_{sp}$, which makes this exotic superconducting
phase potentially realizable in experiments.
With stronger repulsion, the renormalized Luttinger parameter $\gamma$ decreases. Eventually with
repulsion larger than some critical strength, we have
$\gamma<1$, and the repulsive superconducting phase gives way to the ODW phase.

With attractive interaction, the condition $Y/U>0$ gives $Y<0$. Thus
$g_3$ is relevant and flows to $+\infty$ when $C<0$. The sine-Gordon term
$\cos(\sqrt{8\pi} \vartheta_o)$ is locked at $-1$, and  the dominant order
is superconducting SC$^-$, given by
\bea
O_{\text{SC} ^-} &=& \psi{_A} \psibar{_A} - \psi{_B}  \psibar{_B} \nonumber \\
	&\propto & e^{-i\sqrt{2\pi} \vartheta_c} \sin (\sqrt{2\pi} \vartheta_o).
\eea
{In numerical simulations we find the SC$^-$ phase competing with PDW in the strongly
attractive regime. }
When $g_3$ is irrelevant ($C>0$, $Y<0$), the orbital ladder is
in a two component Luttinger liquid phase exhibiting two gapless normal modes and
each mode is a mixture of orbital and charge.

\begin{figure}[htp]
\begin{center}
\includegraphics[angle=0,width=.95\linewidth]{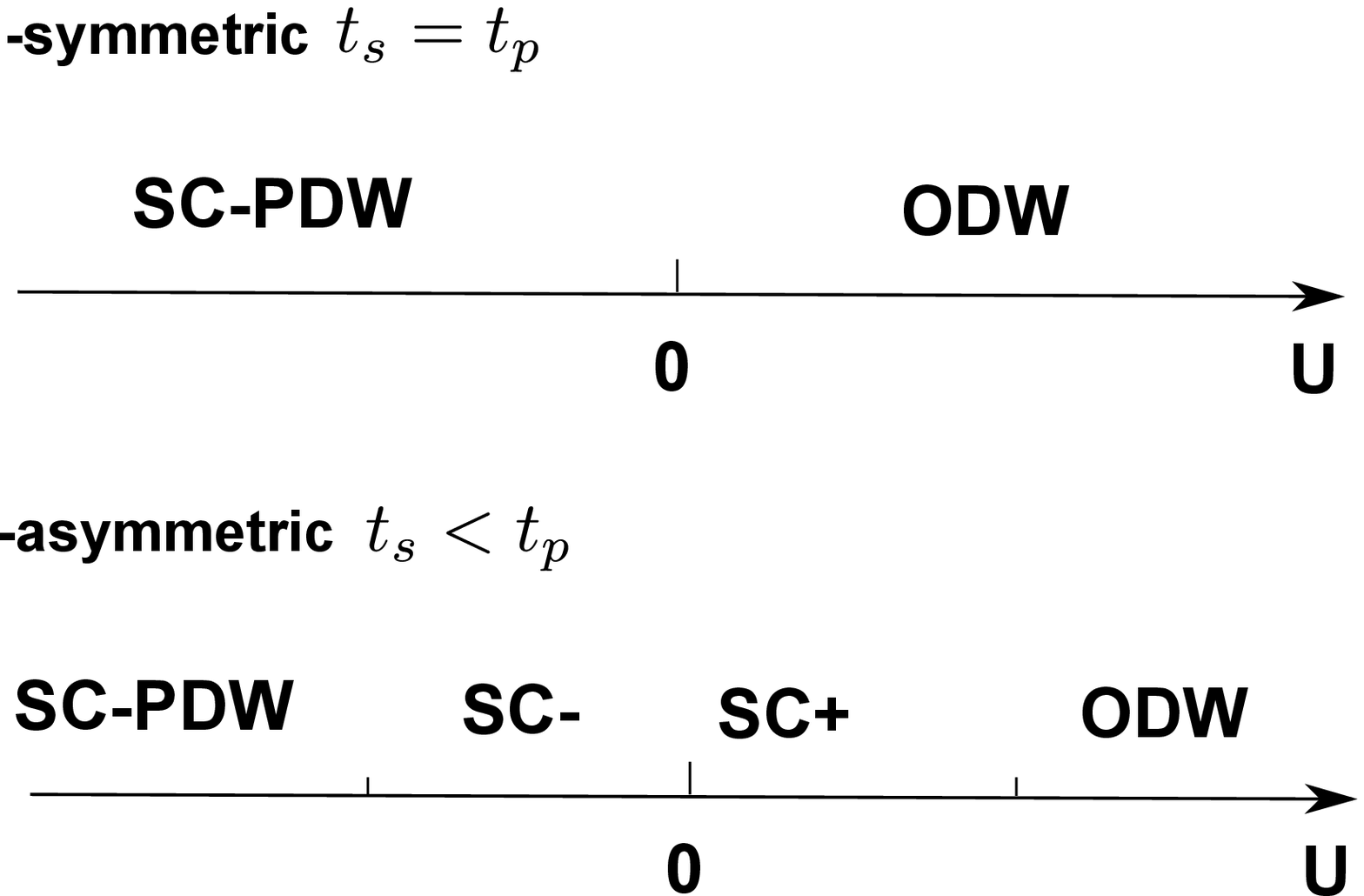}
\end{center}
\caption{{ Schematic phase diagram of the $sp$-orbital ladder. 
In the asymmetric case, the condition that $g_3$ is relevant is taken (see the main text). 
The superconducting SC$^+$ phase appears in the repulsive regime due to quantum fluctuations.  } } 
\label{fig:phasediag}
\end{figure}

{
\paragraph*{Discussion of dipolar tails.---}
In Ref.~\cite{2005_Cirac_dipolartail_PRA}, 
	   it is shown that the tail of dipolar interaction makes correlations 
	   decay as $1/r^3$ in the gapped phase, and that the 
	   tail does not change the critical properties at the critical point. 
	    The present study finds Luttinger liquid phases, which are critical. The power law correlations 
	    characterizing our predicted critical phases decay much slower than $1/r^3$. 
For example, with weak repulsive interaction, the scaling exponent for the SC$^+$ phase, 
$1/\gamma$, is found less than $1$ from Eq.~\eqref{eq:gamma}, well below the dipolar exponent $3$.
	   We thus conclude that the dipolar tail corrections should be negligible.
}
%

\section{Conclusion} 
To conclude, we have studied quantum phases of a one-dimensional $sp$-coupled interacting Fermi gas, 
with both numeric and analytic methods. 
A PDW phase, featuring oscillating Cooper 
pair field, shows up naturally in the attractive regime. 
An incommensurate ODW phase is found in the repulsive regime. A repulsive superconducting phase 
emergent from orbital-charge mixing is also discussed. 
In experiments, radio-frequency spectroscopy can be used to probe
spectra functions~\cite{2011_JILA_RF_Nature,2009_Chen_Levin_RF_PRL},
which exhibit the signatures of pairings of the predicted phases. In orbital
density wave phases, {where there are diverging correlations}
$\langle C_j ^\dag \sigma_y C_j C_{j'} ^\dag \sigma_y C_{j'} \rangle$,
the quench dynamics of occupation numbers of $s$ and $p$ orbitals is a probe
of such orders~\cite{2012_Paramekanti_ProbeCurrent_PRA}.

{\it Acknowledgement.} We appreciate the very helpful discussions with
A. Daley, E. Fradkin and Z. Nussinov, A. Paramekanti and E. Zhao.
This work is supported by A. W. Mellon Fellowship (X.L.),
NSF (PHY11-25915) (X.L.),
AFOSR (FA9550-12-1-0079) (W.V.L.), ARO (W911NF-11-1-0230) and
ARO-DARPA-OLE (W911NF-07-1-0464) (X.L. and W.V.L.), the National Basic
Research Program of China (Grant No 2012CB922101), and Overseas
Scholar Program of NSF of China (11128407) (W.V.L.).

\bibliographystyle{apsrev}
\bibliography{orbitalfermi}
\end{document}